\documentclass[journal]{IEEEtran}

\ifCLASSINFOpdf
\else
   \usepackage[dvips]{graphicx}
\fi
\usepackage{url}
\usepackage{graphicx}
\usepackage{caption}
\usepackage{subcaption}
\usepackage{hyperref}
\usepackage{amsmath}
\usepackage{siunitx}
\usepackage{booktabs}
\usepackage{etoolbox}
\usepackage{tabularx}
\usepackage{makecell}
\usepackage{soul}
\usepackage[T1]{fontenc}
\newrobustcmd{\B}{\bfseries}
\usepackage{cleveref}

\begin{document}

\title{EPG2S: Speech Generation and Speech Enhancement based on Electropalatography and Audio Signals using Multimodal Learning}

\author{Li-Chin Chen, Po-Hsun Chen, Richard Tzong-Han Tsai, and Yu Tsao, \IEEEmembership{Senior Member}
\thanks{This work was supported by the Ministry of Science and Technology in Taiwan under Grant MOST 110-2634-F-008-004- and 109-2634-F-008-006-.}
\thanks{Li-Chin Chen is with the Research Center for Information Technology Innovation, Academia Sinica, 128 Academia Road, Section 2, Nankang, Taipei 115, Taiwan (e-mail: li.chin@citi.sinica.edu.tw).}
\thanks{Po-Hsun Chen is with the Department of Computer Science and Information Engineering, National Central University, No. 300, Zhongda Rd., Zhongli District, Taoyuan City 320317, Taiwan (e-mail: f855189@gmail.com).}
\thanks{Richard Tzong-Han Tsai is with the Department of Computer Science and Information Engineering, National Central University, No. 300, Zhongda Rd., Zhongli District, Taoyuan City 320317, Taiwan (e-mail: thtsai@g.ncu.edu.tw).}
\thanks{Yu Tsao is with the Research Center for Information Technology Innovation, Academia Sinica, 128 Academia Road, Section 2, Nankang, Taipei 115, Taiwan (e-mail: yu.tsao@citi.sinica.edu.tw).}
}

\maketitle

\begin{abstract}
Speech generation and enhancement based on articulatory movements facilitate communication when the scope of verbal communication is absent, e.g., in patients who have lost the ability to speak. Although various techniques have been proposed to this end, electropalatography (EPG), which is a monitoring technique that records contact between the tongue and hard palate during speech, has not been adequately explored. Herein, we propose a novel multimodal EPG-to-speech (EPG2S) system that utilizes EPG and speech signals for speech generation and enhancement. Different fusion strategies based on multiple combinations of EPG and noisy speech signals are examined, and the viability of the proposed method is investigated. Experimental results indicate that EPG2S achieves desirable speech generation outcomes based solely on EPG signals. Further, the addition of noisy speech signals is observed to improve quality and intelligibility. Additionally, EPG2S is observed to achieve high-quality speech enhancement based solely on audio signals, with the addition of EPG signals further improving the performance. The late fusion strategy is deemed to be the most effective approach for simultaneous speech generation and enhancement.
\end{abstract}


\IEEEpeerreviewmaketitle

\section{Introduction}
\IEEEPARstart{S}{peech} signals contain rich acoustic information and are essential for human-human and human-machine communication. However, patients who lose their ability to speak, due to issues such as damage or removal of the vocal cord, poor vocal cord health, and decay in vocal pronunciation, are required to learn alternative ways of communication, e.g., sign language. However, speaking remains the most familiar method of communication. Thus, developing an assisting tool that enables verbal communication based on articulatory movements can be reasonably expected to be beneficial. In addition, as COVID-19 is expected to persist indeterminately, wearing masks has become routine, which obscures oral movements and hinders verbal communication. Thus, techniques capable of converting the movements of the tongue and oral cavity into acoustic signals are relevant in this context. Electropalatography (EPG) is a monitoring technique that records contact between the tongue and hard palate during speech and articulation. EPG has been widely utilized in therapy to improve user articulation. EPG exhibits several advantages over speech signals, e.g., it is unaffected by background noise \cite{EPG, VisualizeEPG} or obstruction of direct vision (e.g., caused by face masks). However, the use of EPG signals to generate or enhance speech has not been adequately researched.

Based on recent advances in machine learning-based technologies, the conversion of biosignals to speech signals has been reported in several studies \cite{Gaddy2020DigitalVO, Chen2021EMA2SAE, Cao2019PermanentMA}. Various signals have been considered for speech generation and enhancement, including surface electromyography (sEMG) \cite{Gaddy2020DigitalVO, Janke2017EMGtoSpeechDG}, electromagnetic articulography (EMA) \cite{Chen2021EMA2SAE, toda2004mapping}, permanent magnetic articulography (PMA) \cite{Cao2019PermanentMA, Gonzlez2016ASS}, ultrasound tongue imaging \cite{Kimura2019SottoVoceAU, Gosztolya2020ApplyingDA}, Doppler signals \cite{Toth2010SynthesizingSF, ma2020noncontact}, visual cues \cite{hou2018audio, chuang2020lite}, and bone-conducted microphone signals \cite{yu2020time}. Further, multimodal learning has been leveraged to integrate information from complementary data, such as text \cite{kinoshita2015text}, videos \cite{hou2018audio}
, bone-conducted microphone signals \cite{yu2020time}, and articulatory movements \cite{Chen2021EMA2SAE}.
However, the transformation of articulatory movements to facilitate communication has not yet been adequately researched.

In real-world application, the artificial palates can be worn as wearable devices to record EPG signals during the translation of articulatory movement into speech. And the verbal speech audio can be recorded before the removal of their vocal cords or during the early stages of decay of pronunciation functionality. In this context, this study intends to explore the possibility of speech generation and enhancement achieved based on EPG and audio speech signals. We propose a novel multimodal EPG-to-speech (EPG2S) system that fully utilizes both EPG and speech signals. Further, combinations of the two modalities and different fusion strategies are discussed. 



\subsection{Speech generation and speech enhancement}
Speech generation involves the artificial production of human speech based on alternative signals. The most commonly used signal is lip reading \cite{Akbari2018Lip2AudspecSR, wand2017improving}. Other speech-related biosignals, such as sEMG \cite{Gaddy2020DigitalVO, Janke2017EMGtoSpeechDG}, EMA\cite{Chen2021EMA2SAE}, PMA \cite{Cao2019PermanentMA, Gonzlez2016ASS}, and ultrasound images \cite{Kimura2019SottoVoceAU, Hueber2010DevelopmentOA}, have also been reported. 
In contrast, speech enhancement is designed to improve speech quality and intelligibility in noisy environments, thereby improving the robustness of the system to environmental noise. 
Notable examples include spectral subtraction \cite{spectralsubtraction1}, the Wiener filter \cite{priori1, prewhitening}, the Karhunen–Loeve transform (KLT) \cite{KLT}, and principal component analysis (PCA) \cite{PCA}. Recently, deep learning (DL)–based models have been adopted in both areas, supporting the direct transformation of bio-signals into speech, and enhancing the quality and intelligibility of speech. 
These models often generate spectral features that are further processed using a vocoder (for example, STRAIGHT \cite{Kawahara1999RestructuringSR}, WORLD \cite{Morise2016WORLDAV}, or a neural network-based vocoder \cite{Yamamoto2020ParallelWA}
) to obtain the synthesized and enhanced speech waveforms. 

\subsection{Multimodal fusion strategies}
The fundamental motivation of multimodal fusion learning is the utilization of complementary information obtained from different modalities to improve performance. The most commonly employed fusion strategies based on combinations of modalities at different stages are early fusion (EF) and late fusion (LF) \cite{fusion}. EF directly concatenates constituent modalities and inputs them into the model. In contrast, LF first processes each signal individually using neural network–based models and then concatenates the output of each network. The concatenated output is processed using another neural network–based model that generates the final speech signals.

\section{Proposed method}

Herein, we propose an EPG2S system for speech generation and enhancement. Furthermore, different combinations of EPG and speech signals using varying fusion strategies are experimented to investigate their differences and effectiveness.


\subsection{Model design}

\begin{figure}
    \centerline{\includegraphics[scale=0.31]{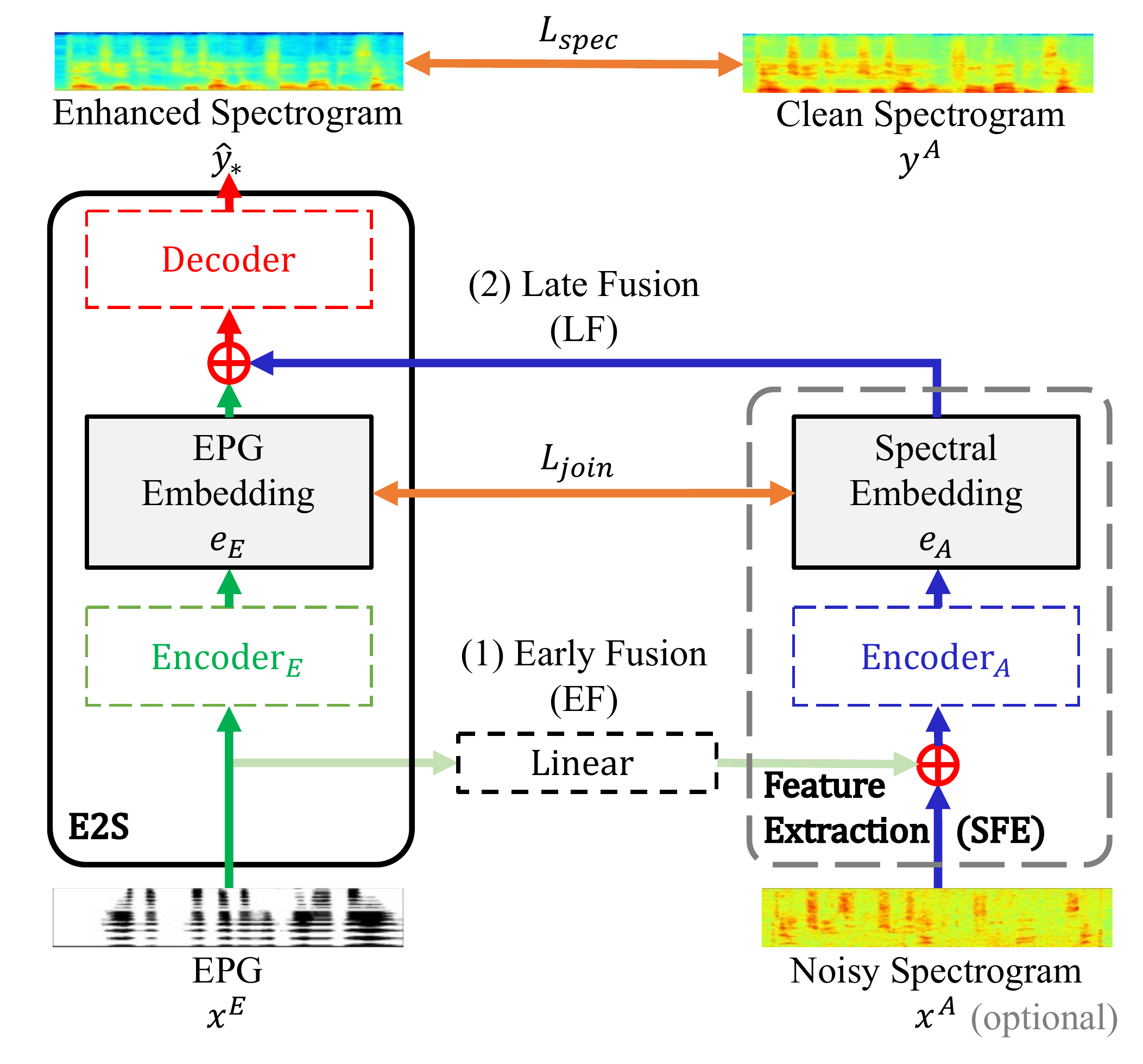}}
    \caption{Overall architecture of the EPG2S system.}
    \label{fig:epg2s_overall}
\end{figure}

\begin{figure}
    \centering
    \begin{subfigure}{.453\columnwidth}
      \centering
      \includegraphics[width=\columnwidth]{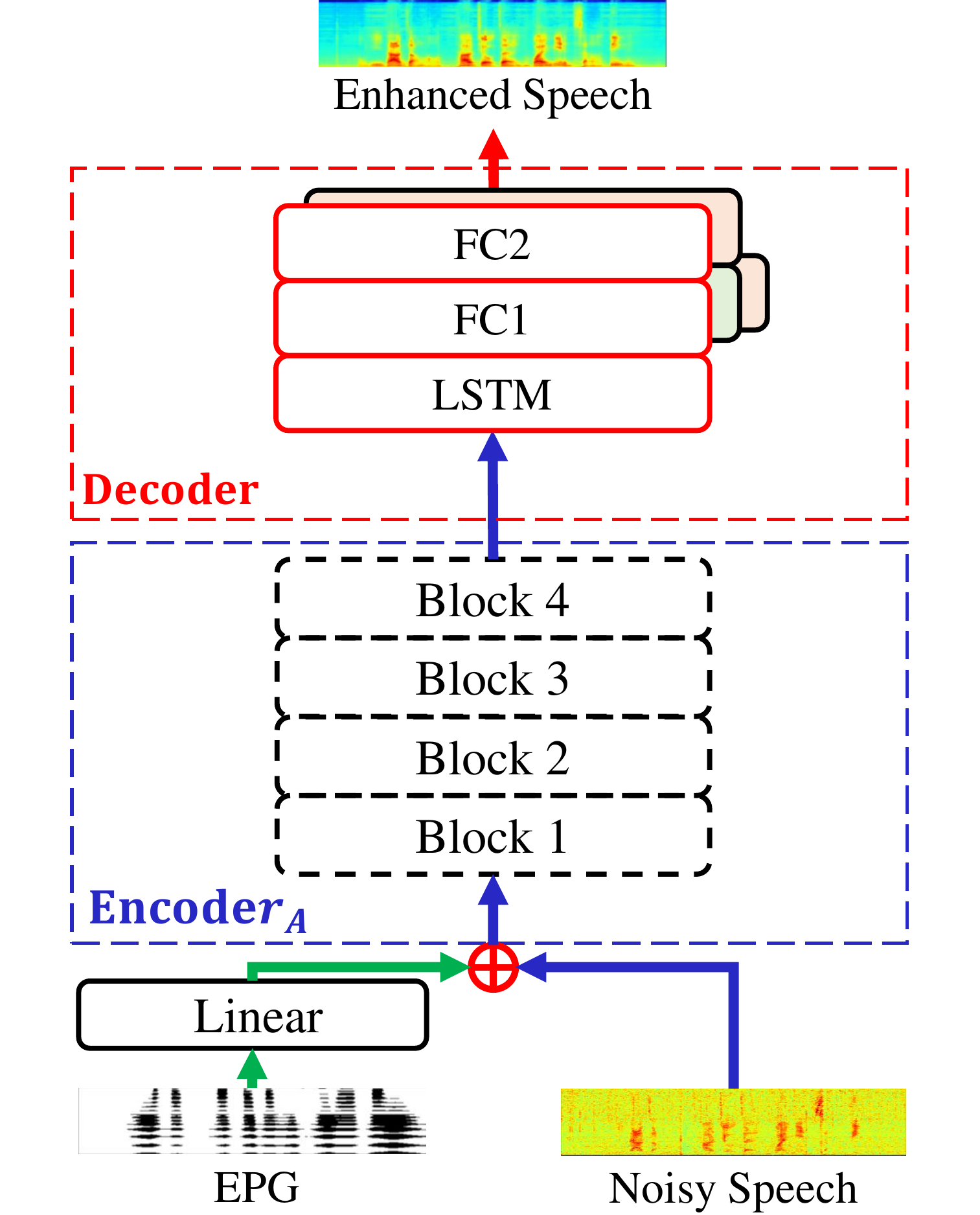}
      \caption{}
    \end{subfigure}%
    \begin{subfigure}{.547\columnwidth}
      \centering
      \includegraphics[width=\columnwidth]{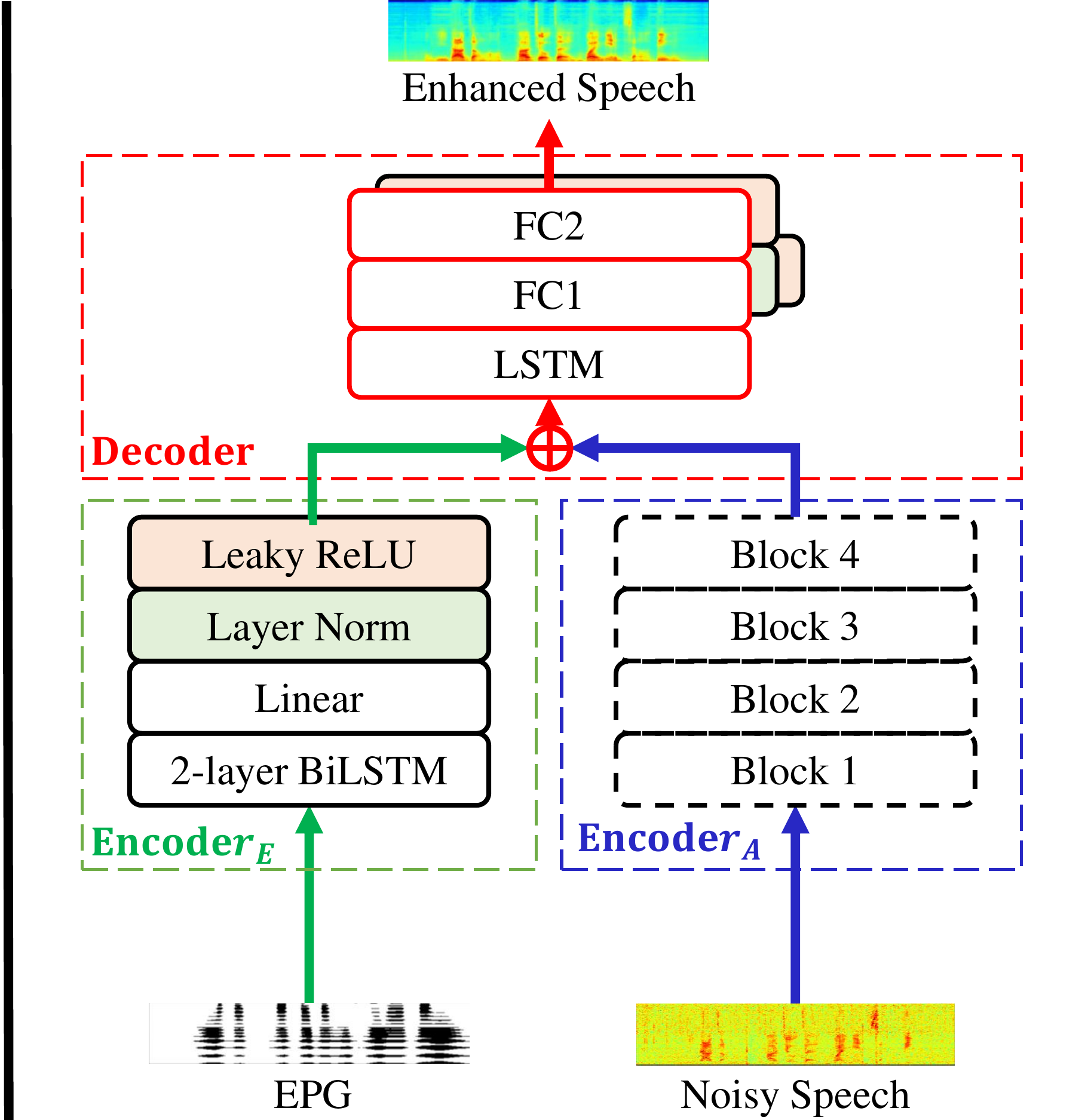}
      \caption{}
    \end{subfigure}
    \caption{EPG2S using different fusion strategies. (a) Using EF strategy, and (b) LF strategy.}
    \label{fig:architecture}
\end{figure}

\begin{figure}
    \centerline{\includegraphics[scale=0.32]{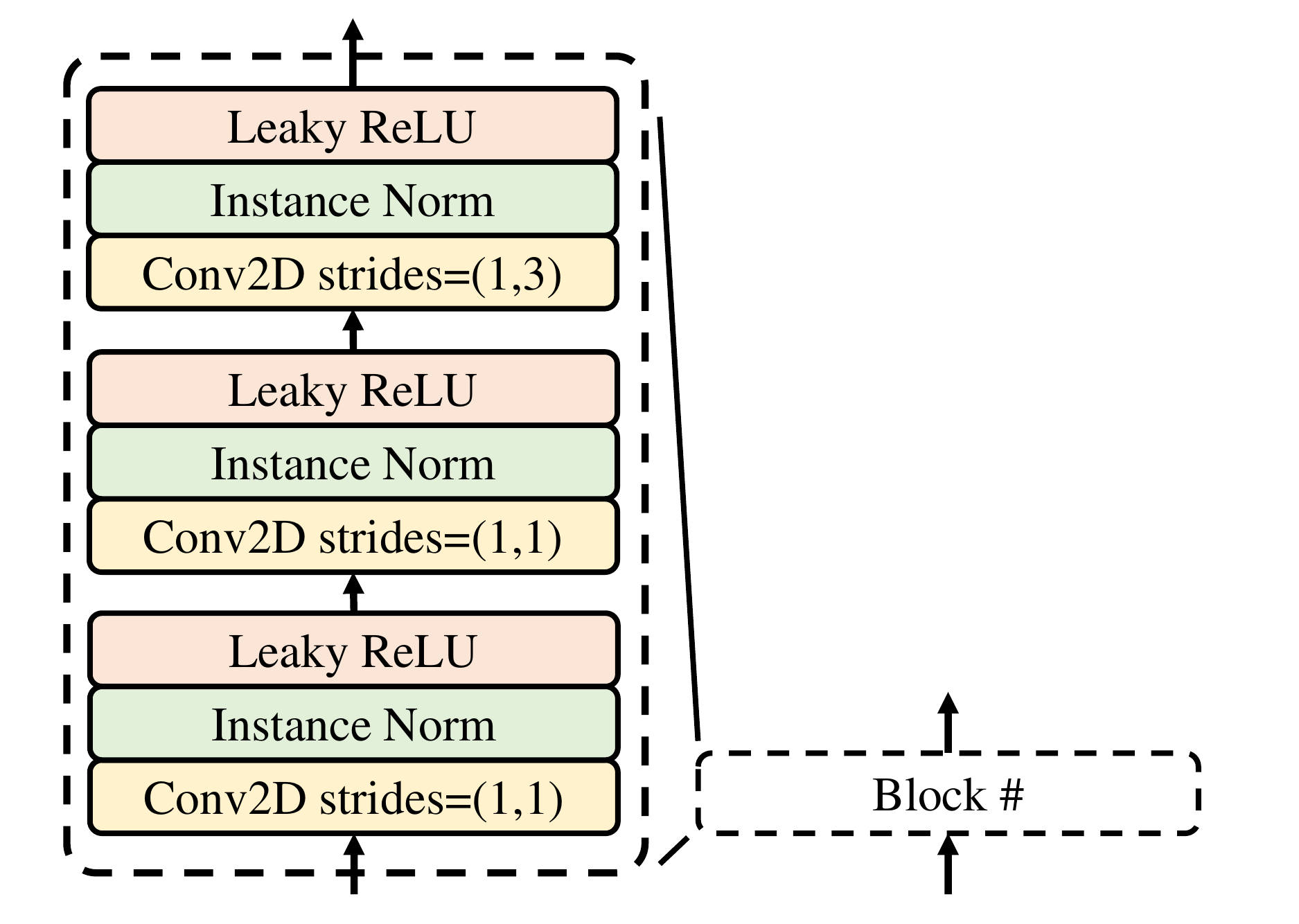}}
    \caption{Structure of the feature extraction component $Encoder_{A}$ for speech signals.}
    \label{fig:Block}
\end{figure}

The overall architecture of EPG2S is depicted in Fig.~\ref{fig:epg2s_overall}, and Fig.~\ref{fig:architecture} presents the structure of (a) EF and (b) LF strategies. EPG2S comprises an EPG-to-speech (E2S) component and a spectrogram feature extraction (SFE) component. E2S includes an $Encoder_{E}$ and a $Decoder$. $Encoder_{E}$ generates a EPG embedding, $e_{E}$. It consists of two layers of bidirectional LSTM (BiLSTM) network, a linear layer, a normalization layer, and a leaky rectified linear activation function (ReLU). The hidden size of the two-layer BiLSTM is 256, and the output size of the linear layer is 512. The $Decoder$ consists of an LSTM layer and two fully connected (FC) layers. The number of hidden nodes in the LSTM is 384, and the output sizes of the two FC layers are 512 and 257. The output obtained from the $Decoder$ is a spectrogram. SFE, the other component, includes $Encoder_{A}$ that uses spectrogram as input and generates a speech embedding, $e_{A}$. $Encoder_{A}$ comprises four blocks. Each block comprises three two-dimensional convolutional (Conv2D) layers, each of which is followed by a normalization layer and a leaky ReLU, as depicted in Fig.~\ref{fig:Block}. The numbers of filters in the first four blocks are 16, 32, 64, and 128, respectively. All kernel sizes are considered to be 3. The final output dimension is 512.

In EF, EPG and speech signals are heterogeneous data types whose dimensions are required to be extended or condensed to achieve concatenation at an early stage. To preserve the rich information in audio files, we extend the dimension of the EPG. The EPG signals are extended via a linear layer and concatenated with the speech spectral vector into a new vector, which becomes the input of the model. The input is then processed using $Encoder_{A}$ and $Decoder$. The EF process can be formulated as follows:

\begin{equation}\label{eq:s_EF}
    \hat y_{EF}[n] = \text{Decoder}\{\text{Encoder}_{A}\{[x^{(A)}[n];\text{Linear}\{x^{(E)}[n]\}]\}\}
\end{equation} where $\hat y_{EF}[n]$ is the generated speech signal.

In contrast, the LF strategy extracts features in advance. $Encoder_{E}$ extracts features from the EPG signal and converts it into a latent representation, $s_{E}[n]$. $Encoder_{A}$ extracts features from speech signals and encodes the spectral features into another latent representations, $s_{A}[n]$. The outputs of $Encoder_{E}$ and $Encoder_{A}$ are then concatenated into an input vector for the fusion layer, $Decoder$. The generated spectrogram is the final output, $\hat y_{LF}[n]$. The LF process is given as follows:

\begin{equation}\label{eq:s_A}
    s_{A}[n] = \text{Encoder}_{A}\{x^{(A)}[n]\},
\end{equation}
\begin{equation}\label{eq:s_E}
    s_{E}[n] = \text{Encoder}_{E}\{x^{(E)}[n]\},
\end{equation}
\begin{equation}\label{eq:s_LF}
    \hat y_{LF}[n] = \text{Decoder}\{s_{A}[n], s_{E}[n]\}.
\end{equation}
where $x^{(E)}[n]$ and $x^{(A)}[n]$ denote the EPG and speech signals, respectively, at the time index, $n$. To obtain better information for spectrogram reconstruction, $Encoder_{A}$ is trained to minimize the differences between the EPG latent representations, $s_{E}[n]$, and corresponding spectral latent representations, $s_{A}[n]$. The loss function is defined as follows:
\begin{equation}\begin{split}\label{eq:EASE2S_loss}
    L_{spec} & = | y^{(A)}[n] - \hat y_{*}[n] | ^2 \\
    L_{join} & = | s_{A}[n] - s_{E}[n] | \\
    L & = L_{spec} + \lambda \cdot L_{join}.
\end{split}\end{equation}
where $y^{(A)}[n]$ denotes the speech signal; $\hat y_{*}$ denotes the output of EPG2S ($\hat y_{LF}$ or $\hat y_{EF}$), $L_{spec}$ denotes the loss of the spectrogram, as measured in terms of the mean square error; $L_{join}$ denotes the deep feature loss of the EPG and spectral latent representations, as measured in terms of the smooth L1 loss; and $\lambda$ denotes a hyperparameter whose value is considered to be 0.1. As both signals might not be obtained simultaneously, EPG2S is trained based on all possible combinations of the two modalities. The input consists of equal proportions of each combination, i.e., 1/3 for each combination (pure EPG signals, pure speech signals, and combined EPG and speech signals). Adam is used as the optimizer, with a learning rate of $1\times10^{-4}$. After generating the spectrogram, the fast Griffin–Lim algorithm \cite{F_GRIFFIN} is used to reconstruct the speech waveforms with the given spectral features.

\section{Experiments}

\subsection{Data collection and pre-processing}
In this study, a CompleteSpeech SmartPalate® system \cite{hardpalate_review} was used to record EPG signals obtained from a male speaker. The speaker read the Taiwan Mandarin hearing in noise test (TMHINT) script \cite{TMHINT}, and the EPG and speech signals were recorded simultaneously in a quiet room. TMHINT is a phone and tone-balanced Taiwanese Mandarin corpus consisting of 320 sentences, with 10 Chinese characters in each sentence.

The EPG sensor included an attachment of 124 electrodes to the upper palate of the speaker. A custom artificial palate was constructed and fitted to their hard palate. Contact between the tongue surface and any of the electrodes (particularly between the lateral margins of the tongue and the borders of the hard palate) triggered the transmission of electronic signals to an external processing unit. Signals were recorded as discrete values of 1 and 0, indicating the presence or absence of contact, respectively. Therefore, the EPG signals were denoted by $s\times124$ vectors, where $s$ denotes time length of the measurements in seconds. The recordings were converted into spreadsheet files, with each column corresponding to a separate electrode. As EPG signals are highly personalized, EPG2S was designed as a personalized system based on individual data. This study was approved by the Institutional Review Board of Acedemia Sinica (Taipei, Taiwan) (AS-IRB-BM-21007).

The speech signals were segmented based on each sentence and paired with the EPG signals. One utterance was excluded because of poor quality. Consequently, 319 pairs of EPG signals and verbal utterances were included in this study. The speech signals were digitalized using a 16-bit analog-to-digital converter at a sampling rate of 16 kHz and stored in waveform audio format. Of the 319 utterances, 222 represented the training set, 27 represented the validation set, and 70 represented the testing set. The short-time Fourier transform was used to convert the speech signals into spectral features. The hop and window sizes were set to 160 and 512, respectively. In addition, the acoustic features were normalized to eliminate magnitude differences between the features. Four types of noise—vehicular, engine, street, and background speaker noise—were artificially added (from 100 noise types \cite{noise100}) to clean speech utterances. 

\subsection{Experiment design}
In this study, we conducted experiments to evaluate the performance of the proposed EPG2S system in speech generation and speech enhancement, respectively. In speech generation, we determined the efficacy of speech generation using EPG, and we further investigated the benefits of adding noisy speech signals during the training stage. In this set of experiments, the combinations used during training included pure EPG signals, EPG and noisy speech signals concatenated using EF, and EPG and noisy speech signals concatenated using LF. The reported results were based on testing using pure EPG signals. 
In the context of speech enhancement, we examined the improvement of distorted speech signals achieved using EPG2S and investigated the feasibility of using EPG signals as auxiliary inputs. First, we investigated the ability of the EPG2S system to enhance speech signals, which was considered the baseline. The model was trained and tested based on distorted noisy speech corresponding to five signal-to-noise-rates (SNRs): -10, -5, 0, 5, and 10 dB. Second, we validated the effectiveness of adding EPG signals to the noisy speech. The performances of EF and LF strategies were compared to demonstrate the effect of incorporating two signals using different approaches.

We evaluated the performance of the proposed EPG2S systems using standardized metrics, including the perceptual evaluation of speech quality (PESQ) \cite{rix2001perceptual}, short-time objective intelligibility (STOI) \cite{STOI}, extended STOI (ESTOI) \cite{ESTOI}, mel cepstral distortion (MCD) \cite{407206,1172250}, and segmental signal-to-noise rate (SSNR) \cite{mermelstein1979evaluation}. PESQ indicates speech quality, and STOI and ESTOI reflect speech intelligibility. 
Higher PESQ, STOI, and ESTOI values correspond to better performance. MCD measures the Euclidean distance between two mel cepstra sequences, which describe the global spectral characteristics of audio signals. In this study, we compared the generated speech with noisy and clean speech to determine their relative distances. Additionally, MCD of enhanced speech was compared to clean speech. SSNR evaluated quantization noise based on the energy in each speech segment. An independent t-test was performed to compare and estimate the degree of difference between the indices of each modality or fusion strategy combination. Validations included comparisons at and across SNR levels. A \emph{p-value} of <0.05 was considered to correspond to statistical significance.

\subsection{Results and discussions}
Table~\ref{tab:evaluation_EPG2S} summarizes the results of speech generation and the t-test. The effects of using different combinations of inputs during training were examined, and all combinations were tested using pure EPG signals. LF outperformed pure EPG in terms of PESQ (\emph{p} < 0.001), ESTOI (\emph{p} < 0.001), and MCD (noisy) (\emph{p} = 0.005) and EF in terms of STOI (\emph{p} < 0.001). Thus, LF can be considered an overall better approach as it achieved better outcomes with respect to a higher number of metrics. The MCD of speech generated using pure EPG was closest to that of clean speech (but not statistically significant), and the corresponding SSNR was significantly higher than those achieved using the other two approaches (\emph{p} < 0.001). The results indicate that the utilization of pure EPG signals led to desirable outcomes. Further, the addition of noisy speech signals improved the quality and intelligibility of the generated speech.


\begin{table}[!t]
\scriptsize
    \caption{\scriptsize Performance in speech generation. (indices and \emph{p-values})}
    \label{tab:evaluation_EPG2S}
    \centering
    
    \sisetup{detect-weight,mode=text}
    \renewrobustcmd{\bfseries}{\fontseries{b}\selectfont}
    \renewrobustcmd{\boldmath}{}
    \addtolength{\tabcolsep}{-1.0pt}

    \begin{tabularx}{\columnwidth}{p{18mm}p{20pt}p{20pt}p{20pt}p{22pt}p{22pt}p{22pt}}
    \toprule
    Training setting & PESQ  & STOI  & ESTOI & MCD(N) & MCD(C) & SSNR \\
    \midrule
    Pure EPG & 2.039 & 0.583 & 0.321 & 10.287 & \B 5.173 & \B -3.628\\
    
    EPG + N (EF) & 1.996 & 0.588 & \B 0.346 & 10.380 & 5.389 & -4.088\\
    EPG + N (LF) & \B 2.048 & \B 0.592 & 0.341 & \B 10.483 &5.311 & -3.881\\
    \midrule
    Pure EPG - EF & 0.000* & 0.001* & 0.058 & 0.163 & 0.393  & 0.000*\\
    Pure EPG - LF & 0.000* & 0.741 & 0.000* & 0.005* & 1.000 & 0.000*\\
    EF - LF & 0.805 & 0.000* & 0.002* & 0.160 & 0.393 & 0.000*\\
    \bottomrule
    \multicolumn{7}{p{250pt}}{\scriptsize Model trained with different combination of input. All were tested with pure EPG signals. Input combinations include pure EPG signals; EPG + noisy speech using EF; and EPG + noisy speech using LF. N: noisy speech; C: clean speech; *: \emph{p-value} < 0.05.}
    \end{tabularx}
\end{table}

\crefformat{footnote}{#2\footnotemark[#1]#3}

Tables ~\ref{tab:evaluation_baseline} and ~\ref{tab:evaluation_fusion} summarize the results of speech enhancement. Table~\ref{tab:evaluation_baseline} presents a comparison of the differences between unprocessed noisy signals and EPG2S-processed speech corresponding to different SNRs. The results indicate that EPG2S is effective in speech enhancement based on pure audio signals. All metrics improved significantly across SNRs (\emph{p} < 0.001), except for SSNR (\emph{p} = 0.795). STOI and SSR at 10 dB and SSR at 5 dB did not improve when processed with audio signals only. Table~\ref{tab:evaluation_fusion} summarizes the results obtained using the EPG2S trained and tested on EPG and speech signals combined using different fusion strategies corresponding to different SNRs. The results indicated that, compared to the baseline, EPG signals were beneficial as an auxiliary input during speech enhancement, irrespective of the fusion strategy (both EF and LF exhibited \emph{p} < 0.001 corresponding to all metrics compared to the baseline)\footnote{\label{note1}\url{https://github.com/ishiou/EPG2S/blob/main/p_value_table.md}}. The LF strategy performed better than EF in terms of MCD (\emph{p} < 0.001) and SSNR (\emph{p} < 0.001). Although EF and LF appeared to perform better corresponding to different SNRs, they did not attain statistical significance. The results indicated that the availability of both EPG and speech signals improved the performance of EPG2S. Complete statistical results\cref{note1} and audio samples\footnote{\url{https://ishiou.github.io/EPG2S/EPG2S_audio_example.html}} are available on GitHub.

\begin{table}[!t]
\scriptsize
    \caption{\scriptsize Comparison of unprocessed noisy speech and EPG2S processed noisy speech (without EPG) under different SNRs in speech enhancement.}
    \label{tab:evaluation_baseline}
    \sisetup{detect-weight,mode=text}
    \renewrobustcmd{\bfseries}{\fontseries{b}\selectfont}
    \renewrobustcmd{\boldmath}{}
    \addtolength{\tabcolsep}{-1.0pt}
    \newcolumntype{Y}{>{\centering\arraybackslash}X}
    \begin{tabularx}{\columnwidth}{r p{10pt}p{10pt}p{12pt}p{12pt}p{22pt}p{10pt}p{10pt}p{12pt}p{12pt}p{22pt}}
    \toprule
            & \multicolumn{5}{c}{Unprocessed Noisy Speech}
            & \multicolumn{5}{c}{EPG2S Processed Speech (baseline)}
            \\
    \cmidrule(lr){2-6} \cmidrule(lr){7-11}
            & PESQ  & STOI  & ESTOI & MCD & SSNR
            & PESQ$^{\dagger}$  & STOI$^{\dagger}$  & ESTOI$^{\dagger}$ & MCD$^{\dagger}$ & SSNR \\
    \midrule
    10 dB   & 1.838 & \B 0.823 & 0.615 & 7.342 & \B 0.157*
            & \B 2.812* & 0.816 & \B 0.654* & \B 5.399* & -4.054 \\
    
    5 dB    & 2.565 & 0.756 & 0.493 & 8.762 & \B -2.570*
            & \B 3.100* & \B 0.829* & \B 0.647* & \B 5.628* & -4.251\\
            
    0 dB    & 2.308 & 0.668 & 0.365 & 10.154 & -4.918
            & \B 2.841* & \B 0.783* & \B 0.560* & \B 5.809* & \B -4.438*\\
            
    -5 dB   & 2.035 & 0.564 & 0.249 & 11.493 & -6.928
            & \B 2.503* & \B 0.704* & \B 0.437* & \B 6.137* & \B -4.764*\\
            
    -10 dB  & 1.779 & 0.469 & 0.158 & 12.723 & -8.441
            & \B 2.078* & \B 0.578* & \B 0.284* & \B 6.437* &  \B -5.069* \\
    \midrule
    Avg.    & 2.105 & 0.656 & 0.376 & 10.095 & -4.540
            & \B 2.667 & \B 0.742 & \B 0.516 & \B 5.882 &  \B -4.515 \\

    \bottomrule
    \multicolumn{11}{p{250pt}}{\scriptsize *: \emph{p-value} < 0.05 at SNRs; $^{\dagger}$: \emph{p-value} < 0.05 across SNRs.}
    \end{tabularx}
\end{table}

\begin{table}[!t]
\scriptsize
    \caption{\scriptsize Comparison of using different fusion strategies to combine EPG and noisy speech signals cooresponding to different SNRs in speech enhancement.}
    \label{tab:evaluation_fusion}
    
    \sisetup{detect-weight,mode=text}
    \renewrobustcmd{\bfseries}{\fontseries{b}\selectfont}
    \renewrobustcmd{\boldmath}{}
    \addtolength{\tabcolsep}{-1.0pt}
    \newcolumntype{Y}{>{\centering\arraybackslash}X}
    \begin{tabularx}{\columnwidth}{r p{10pt}p{10pt}p{12pt}p{12pt}p{20pt}p{10pt}p{10pt}p{12pt}p{12pt}p{22pt}}
    \toprule
            & \multicolumn{5}{c}{EPG2S (EF)}
            & \multicolumn{5}{c}{EPG2S (LF)}
             \\
    \cmidrule(lr){2-6} \cmidrule(lr){7-11}
            & PESQ  & STOI  & ESTOI & MCD & SSNR
            & PESQ  & STOI  & ESTOI & MCD$^{\dagger}$ & SSNR$^{\dagger}$ \\
    \midrule
    10 dB   & \B 2.813* & \B 0.820 & \B 0.660 & 5.221 & -3.572
            & 2.732 & 0.813 & 0.648 & \B 5.153 & \B -3.382*\\
    
    5 dB    & \B 3.160* & \B 0.836 & \B 0.662 & 5.361 & -3.771
            & 3.146 & 0.830 & 0.657 & \B 5.182* & \B -3.540* \\
            
    0 dB    & 2.916 & \B 0.795 & 0.585 & 5.528 & -4.003
            & \B 2.935 & 0.791 & \B 0.588 & \B 5.292* & \B -3.717* \\
            
    -5 dB   & 2.638 & 0.735 & 0.494 & 5.783 & -4.243
            & \B 2.688 & \B 0.738 & \B 0.510 & \B 5.458* & \B -3.936*\\
            
    -10 dB  & 2.311 & 0.648 & 0.391 & 5.986 & -4.483
            & \B 2.427 & \B 0.676* & \B 0.430* & \B 5.614* & \B -4.208*\\
    \midrule
    Avg.    & 2.768 & 0.767 & 0.558 & 5.576 & -4.014
            & \B 2.786 & \B 0.770 & \B 0.567 & \B 5.340 & \B -3.757\\
    \bottomrule
     \multicolumn{11}{p{250pt}}{\scriptsize *: \emph{p-value} < 0.05 at SNRs; $^{\dagger}$: \emph{p-value} < 0.05 across SNRs.}
    \end{tabularx}
\end{table}

\section{Conclusions}
In this study, an EPG2S system was proposed for speech generation and enhancement based on multimodal learning. To the best of our knowledge, no previous study has applied EPG signals to speech generation and enhancement. Our results demonstrated that pure EPG signals can be used to achieve desirable outcomes in speech generation. Further, the incorporation of additional noisy speech signals improved quality and intelligibility. EPG also accelerated speech enhancement performance. 
LF performed better than EF in speech generation and enhancement. We successfully constructed a personalized system based on personalized signals. Our observations indicated that patient communication can be adequately supported using EPG signals. Audio recording is optional, nevertheless, beneficial if obtainable. However, the construction of personalized models for multiple individuals remains to be fully verified, which is a limitation of this study. Further investigations are required before this technology can be widely applied in practice.


\bibliographystyle{IEEEbib}
{\footnotesize\bibliography{refs}}

\end{document}